\documentclass[printer]{aa}
\usepackage{graphics,epsfig,lscape,txfonts}
\usepackage{natbib,aas_macros}
\bibpunct{(}{)}{;}{a}{}{,}

\def\cm-2{cm$^{-2}$}

\def\ein{{\it Einstein}}
\def\chandra{{\it Chandra}}
\def\sax{{{\it Beppo}SAX}}
\def\xmm{{XMM-Newton}}
\def\n253{\object{NGC~253}}
\def\m33{\object{M33}}
\def\mx7{\object{M33 X$-$7}}
\def\x7{\hbox{X$-$7}}

\newcommand{\ergs}[1]{$\times10^{#1}$ \hbox{erg s$^{-1}$}}
\newcommand{\oergs}[1]{$10^{#1}$ erg s$^{-1}$}
\newcommand{\hcm}[1]{$\times10^{#1}$ cm$^{-2}$}

\newcommand{\nh}{\hbox{$N_{\rm H}$}}

\begin{document}

   \title{The eclipsing massive X-ray binary M33 X$-$7:
    New X-ray observations and optical identification\thanks{This work is based 
    on observations obtained with XMM-Newton, an ESA Science Mission 
    with instruments and contributions directly funded by ESA Member
    States and the USA (NASA).}}

   \author{W.~Pietsch\inst{1} \and 
	   B.~J.~Mochejska \inst{2,3} \and
	   Z.~Misanovic\inst{1} \and
	   F.~Haberl\inst{1} \and
	   M.~Ehle\inst{4} \and
	   G.~Trinchieri\inst{5}
          }
\institute{Max-Planck-Institut f\"ur extraterrestrische Physik, 85741 Garching, Germany 
          \and Harvard-Smithsonian Center for Astrophysics, 60 Garden Street,
	  Cambridge, MA 02138, US
	  \and Hubble Fellow
	  \and \xmm\ Science Operations Centre, ESA, Villafranca del Castillo,
	  P.O. Box 50727, 28080 Madrid, Spain
	  \and Osservatorio Astronomico di Brera, via Brera 28, 20121 Milano, Italy  }
     
     \offprints{W.~Pietsch}
     \mail{wnp@mpe.mpg.de}

   \date{Received; accepted }
   \titlerunning{The eclipsing massive X-ray binary M33 X$-$7}

	\abstract{The eclipsing X-ray binary \mx7\ was in the field of view
	during  several observations of our \xmm\ \m33\ survey and in the
	archival \chandra\  observation 1730 which cover a large part of 
	the 3.45 d orbital
	period. We detect emission of \mx7\ during eclipse and a soft X-ray
	spectrum of the source out  of eclipse that can best be described by
	bremsstrahlung or disk blackbody models. No significant regular
	pulsations of the source in the range 0.25--1000~s were found. The
	average source luminosity out of eclipse is 5\ergs{37} (0.5--4.5 keV).
	In a special analysis of DIRECT observations we identify as optical
	counterpart a B0I to O7I star of 18.89 mag in V which
	shows the ellipsoidal heating light curve of a high mass X-ray binary
	with the \mx7\ binary period. The location of the X-ray eclipse and the
	optical minima allow us to determine an improved binary period and 
	ephemeris of mid-eclipse as
	HJD~$(245\,1760.61\pm0.09)\pm N\times(3.45376\pm0.00021)$. The mass of
	the compact object derived from orbital parameters and the optical 
	companion mass, the lack of pulsations, and the
	X-ray spectrum of \mx7\ may indicate that the compact object in the system
	is a black hole. \mx7\ would be the first detected eclipsing high mass black
	hole X-ray binary. 
	
} 
\maketitle

\section{Introduction}
\mx7 (hereafter \x7) 
was detected as a variable source with a luminosity brighter than 
\oergs{38} in \ein\ observations 
\citep{1981ApJ...246L..61L,1983ApJ...275..571M,
1988ApJ...325..531T,1988ApJ...329.1037T}. \citet{1989ApJ...336..140P} suggested
that the \x7 variability pattern can be explained by an eclipsing X-ray binary
(XRB) with an orbital period of 1.7 d and an eclipse duration of $\sim0.4$ d.
This finding was the first identification of a close accreting binary system
with an X-ray source in an external galaxy other than the Magellanic Clouds. It
was confirmed combining \ein\ observatory and first ROSAT data 
\citep{1993ApJ...418L..67S,1994ApJ...426L..55S}. 
With the inclusion of more ROSAT and ASCA data 
\defcitealias{1999MNRAS.302..731D}{D99}
\citep[][ hereafter \citetalias{1999MNRAS.302..731D}]{1997AJ....113..618L,1999MNRAS.302..731D} 
the orbital period turned out to be twice as long.
The shape of the eclipse could be described by a slow ingress 
($\Delta \Phi_{\rm ingress} = 0.10\pm0.05$), 
an eclipse duration of  $\Delta \Phi_{\rm eclipse} = 0.20\pm0.03$, and a fast
eclipse egress ($\Delta \Phi_{\rm egress} = 0.01\pm0.01$) with an ephemeris for
the mid-eclipse time of HJD~244\,8631.5$\pm$0.1 + N$\times$(3.4535$\pm$0.0005).  
In addition, \citetalias{1999MNRAS.302..731D} discovered evidence for a 0.31~s pulse period.
The orbital period, pulse period and observed X-ray luminosity are remarkably
similar to those of the Small Magellanic Cloud neutron star XRB \object{SMC X$-$1} 
\citep{2000A&AS..147...25L}. However, if the pulse period of \x7 can not be
confirmed, the source could also resemble high mass black hole XRBs (BHXB) like
\object{LMC X$-$1} or \object{LMC X$-$3}. It would be the first eclipsing object within
this rare class of XRBs.

The position of \x7 correlates with the dense  O--B association HS13 
\citep{1980APJS...44..319H} and therefore no individual counterpart could be 
identified based on position only. However, its location in HS13
is consistent with the expectation of a massive companion. As \citetalias{1999MNRAS.302..731D} point out, 
the optical counterpart is likely to show ellipsoidal and/or X-ray 
heating variations \citep{1986A&A...154...77T} which 
can be used for the optical identification.

Variable optical sources within \m33 were systematically searched for in the 
DIRECT project 
\defcitealias{2001AJ....122.2477M}{M01b}
\citep[see e.g.][ hereafter \citetalias{2001AJ....122.2477M}]{2001AJ....122.2477M}. Many eclipsing
binaries, Cepheids, and other periodic, possibly long-period or nonperiodic
variables were detected. \x7 is located in 
DIRECT field M33B. The variability of the optical counterpart was not detected in the
previous analysis due to the limitations of the variable search
strategy for such small amplitude variables in crowded regions.

As a follow-up of our study of the X-ray source population of \m33 based on all
archival ROSAT observations 
\defcitealias{2001A&A...373..438H}{HP01}
\citep[][ hereafter \citetalias{2001A&A...373..438H}]{2001A&A...373..438H}, 
we planned a deep
\xmm\ raster survey of \m33\ based on 22 Telescope Scientist guaranteed time 
(proposal no 010264) and AO2 (proposal no 014198) observations, each with a
duration of about 10 ks
\citep[for first results see][]{2003AN....324...85P}. 
\x7 was covered in 13 of these observations at varying off-axis angles 
and covering different orbital phases. 

In this paper we report on
time and spectral variability of \x7 within the \xmm\ raster survey. We add
results from an archival \chandra\ observation, which covered the source, and 
a dedicated timing analysis of the DIRECT data of the HS13 region. 

\section{X-ray observations and results}
\begin{table*}
\begin{center}
\caption[]{\mx7 observations with the observatories \xmm\
(proposal numbers 010264 and 014198) and \chandra\ (1730). Besides observation
0102642101, where \mx7 was only in the field of view of the MOS detectors, for 
\xmm\ we give EPIC PN count rates, hardness ratios and luminosities. For the low
state observations, no hardness ratio could be determined.}
\begin{tabular}{llrrrrrcrc}
\hline\noalign{\smallskip}
\hline\noalign{\smallskip}
\multicolumn{1}{c}{Obs. id.} & \multicolumn{1}{c}{Obs. dates} &
\multicolumn{1}{c}{Elapse time} & \multicolumn{1}{c}{$R_{\rm ext}$} & 
\multicolumn{1}{c}{Count rate$^{**}$} & \multicolumn{1}{c}{HR} &
\multicolumn{1}{c}{$L_{\rm X}^{**}$} & \multicolumn{1}{c}{Offax}& 
\multicolumn{1}{c}{Binary phase} \\ 
\noalign{\smallskip}
& & \multicolumn{1}{c}{(ks)} & \multicolumn{1}{c}{(\arcsec)}&  
\multicolumn{1}{c}{(ct ks$^{-1}$)} & & (\oergs{37})& 
\multicolumn{1}{c}{(\arcmin)}\\
\noalign{\smallskip}
\multicolumn{1}{c}{(1)} & \multicolumn{1}{c}{(2)} & \multicolumn{1}{c}{(3)} & 
\multicolumn{1}{c}{(4)} & \multicolumn{1}{c}{(5)} & \multicolumn{1}{c}{(6)} & 
\multicolumn{1}{c}{(7)} & \multicolumn{1}{c}{(8)} & \multicolumn{1}{c}{(9)} \\
\noalign{\smallskip}\hline\noalign{\smallskip}
0102641201 & 2000-08-02 & 13.1~~ & 30.0 & $172\pm~~~4$ & $1.40\pm0.38^{\dagger}$ 
& 8.7 & 10.4 &0.54--0.58 \\
0102640401 & 2000-08-02 & 13.1~~ & 40.0 & $163\pm~~~5$ & $1.27\pm0.23^{\dagger}$ 
& 8.5 & 10.4 &0.61--0.66\\
0102640101 & 2000-08-04 & 13.3~~ & 22.5 & $4.0\pm1.2$ & $$ 
& 0.2 & 8.3 &0.11--0.16\\
0102640501 & 2001-07-05 & 11.8~~ & 45.0 & $57\pm~~~2$ & $1.27\pm0.30~~$ 
& 8.3 & 7.1 &0.33--0.36\\
0102640601 & 2001-07-05 & $8.2^{+}$ & 35.0 & $75\pm~~~5$ & $1.09\pm0.45~~$ 
& 6.0 & 15.7 &0.23--0.26\\
0102640701 & 2001-07-05 & 11.7~~ & 35.0 & $54\pm~~~2$ & $1.10\pm0.26~~$ 
& 7.8 & 9.3 &0.28--0.31\\
0102541101 & 2001-07-08 & 12.5~~ & 30.0 & $5.1\pm1.7$ & $$ 
& 0.3 & 11.1 &0.01--0.05\\
0102642101 & 2002-01-25 & $12.3^{*}$ & 35.0 & $115\pm~~3^{*}$ & $1.61\pm0.35^{*}$ 
& 8.5 & 13.1 &0.25--0.29\\
0102642301 & 2002-01-27 & 12.3~~ & 35.0 & $316\pm~~~6$ & $1.17\pm0.20~~$ 
& 7.0 & 0.9 &0.81--0.85\\
0141980501 & 2003-01-22 &  8.1~~ & 35.0 & $114\pm~10$ & $1.20\pm0.15~~$ 
& 9.0 & 8.3 &0.18--0.21\\
0141980601 & 2003-01-23 & 13.6~~ & 37.5 & $153\pm~~~4$ & $1.23\pm0.28~~$ 
& 7.3 & 10.4 &0.46--0.50\\
0141980701 & 2003-01-24 & 13.7~~ & 35.0 & $164\pm~~~6$ & $1.12\pm0.34~~$ 
& 7.3 & 10.4 &0.52--0.56\\
0141980801 & 2003-02-12 & 10.2~~ & 35.0 & $221\pm~~~6$ & $1.16\pm0.28~~$ 
& 8.6 & 8.3 &0.20--0.23\\
1730       & 2000-12-07 & 52.0~~ & 15.0 & $100\pm~~~2$ & $1.89\pm0.09~~$ 
& 6.3 & 8.6 &0.49--0.65\\
\noalign{\smallskip}
\hline
\noalign{\smallskip}
\end{tabular}
\label{observations}
\end{center}
Notes and references:\\
$^{ +~}$: \mx7 only in PN field of view. \\
$^{ *~}$: \mx7 only in MOS field of view, integrated MOS count rate, hardness
ratio and luminosity given. \\
$^{ {\dagger}~}$: The on average higher HR during
observations 0102641201 and 0102640401 compared to the other observations 
reflects the difference between
thick and medium filter and most likely not a change of the spectrum of \mx7.\\
$^{ **}$: raw count rate and luminosity in the 0.5--4.5 keV band assuming the 
best fitting thermal bremsstrahlung spectrum (see Table~\ref{epic_spectra}) corrected for a Galactic 
foreground absorption of \nh = 6.38\hcm{20} \citep{1990ARA&A..28..215D} 
and a distance of \m33 of
795 kpc \citep{1991PASP..103..609V},  which we use throughout the paper.\\
\end{table*}

For the detailed analysis of \x7\ we mostly used data from \xmm\ EPIC. From the
\m33 observations in the \chandra\ archive only the observation with identification
no 1730 covered the \x7\ field.
Table~\ref{observations} summarizes the observation
identifications (Col. 1), observation dates (2), elapse time (3), 
extraction radius $R_{\rm ext}$ used for count rates, light curves, and 
spectra (4),
\x7 raw count rates in the 0.5--4.5 keV band (5), hardness ratios (6), 
and luminosities in the 0.5--4.5 keV band (7). For the \xmm\ observations, 
values in columns 4 
to 7 correspond to the EPIC PN detector if not indicated differently, since it
gives about twice the number of photons than the EPIC MOS cameras.
Off axis angle of \x7 (8) and phase coverage within the binary orbit according 
to the ephemeris of
\citetalias{1999MNRAS.302..731D} (9) are also given. As hardness ratio (HR) we use the
ratio of the counts in the 1.2--3.0 keV band to the counts in the 0.5--1.2 keV
band. 
Luminosities were determined from thermal bremsstrahlung spectra
(see Sect. 2.3). During observations  0102640101 and  0102641101 \x7 was
in low state and source statistics did not allow us to derive hardness
ratios. For these observations we assumed that the source spectrum was similar
to the spectrum during observation 0102642301 and scaled the source 
luminosity from the vignetting corrected count rates. 

In the \xmm\ observations \citep{2001A&A...365L...1J} the EPIC PN and MOS instruments 
\citep{2001A&A...365L..18S,2001A&A...365L..27T} were mostly operated in the 
full frame mode resulting in  a time
resolution of 73.4 ms and 2.6 s, respectively. Only for the first two observations
in Table~\ref{observations}, the PN detector was operated in the extended full
frame mode (time resolution 200 ms) and during observation 0102640101 the MOS
detectors were operated in the small window mode (0.3~s time resolution for the
inner CCDs).
The medium filter was in front of the EPIC cameras in all but the first two 
observations which were performed with the thick filter. 
We used all EPIC instruments for imaging, position determination
and for the timing and spectral investigations  of \x7. In most of the
observations the source is located at high off-axis angle (see 
Table~\ref{observations}), and could be outside of the field of view in some of 
the cameras. Also, the cameras normally cover different times. 
The \xmm\ point spread function (PSF) required extraction radii  $R_{\rm ext}$ 
larger than 22\farcs5 to encircle $>80\%$ of the source photons. 
Depending on the location of the source, counts could be missing due to CCD
gaps. Many of the \xmm\ 
observations suffer from times of high particle background.  To be able to
also use these times to cover as much as possible of the \x7 binary orbit, we 
restricted the energy band for light curve and hardness ratio analysis to
0.5--3.0 keV where the source is brightest.

Four \chandra\ ACIS observations \citep{2000SPIE.4012....2W} of \m33\ were obtained from the \chandra\ 
Data Archive (http://asc.harvard.edu/cgi-gen/cda). However, only the ACIS I 
observation 1730 (see Table~\ref{observations}) covered the \x7 field.
The instrument was operated in the full frame
mode (3.2 s time resolution). \x7\ is positioned in the outer corner of the 
front-illuminated CCD chip I2 during the observation. 

The deep space orbits of the satellites \xmm\ and \chandra\ led to
long continuous observation times of \x7. The low earth orbits of 
the \ein, ROSAT, ASCA and \sax\ observatories on the other hand, led to observations split in
many short intervals of typically less than 1\,500~s. 

The data analysis was performed using tools in the SAS v5.4.0, CIAO v2.3,
EXSAS/MIDAS 1.2/1.4, and 
FTOOLS v5.2 software packages, the imaging application DS9 v2.1b4 , the timing 
analysis package XRONOS v5.19 and spectral analysis software XSPEC v11.2.   

For the time variability investigations all \x7\ event times were 
corrected to solar system barycenter arrival times.

\subsection{Time variability}
\begin{figure}
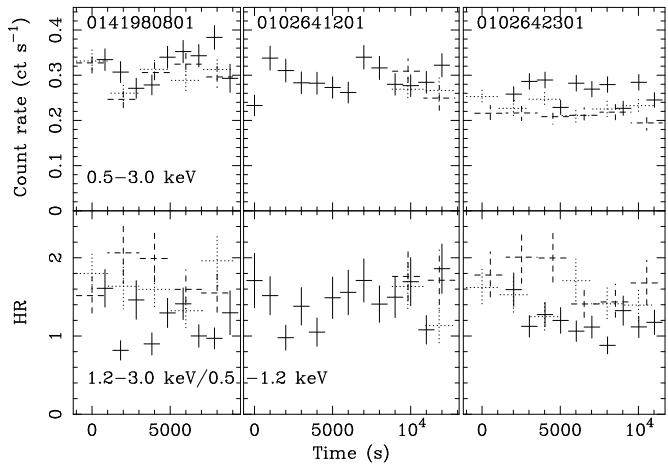

   \resizebox{\hsize}{!}{
   \includegraphics[bb=72 25 546 267,angle=-90,clip]{0081f1a.ps}
   \includegraphics[bb=72 93 546 317,angle=-90,clip]{0081f1b.ps}
   \includegraphics[bb=72 92 546 304,angle=-90,clip]{0081f1c.ps}}
    \caption[]{\xmm\ EPIC light curves and hardness ratio of \mx7 during observations 
    0141980801, 0102641201, and 102642301 with time zero corresponding to 
    HJD 245\,2683.14827, 245\,1758.79473, 245\,2301.91542, respectively (solar 
    system barycenter corrected). Count rates were corrected for vignetting.
    EPIC MOS rates were scaled by a factor of 2.5 to approximately correct for 
    the difference in instrument efficiency compared to EPIC PN. EPIC PN, MOS1
    and MOS2 data are marked with solid, dotted, and dashed error bars. The
    light curves cover orbital phases 0.20--0.23, 0.54--0.58, and 0.81--0.85
    (left to right, ephemeris of \citetalias{1999MNRAS.302..731D}).
    \label{epic_lc}}
\end{figure}
\begin{figure}
   \resizebox{\hsize}{!}{\includegraphics[angle=-90]{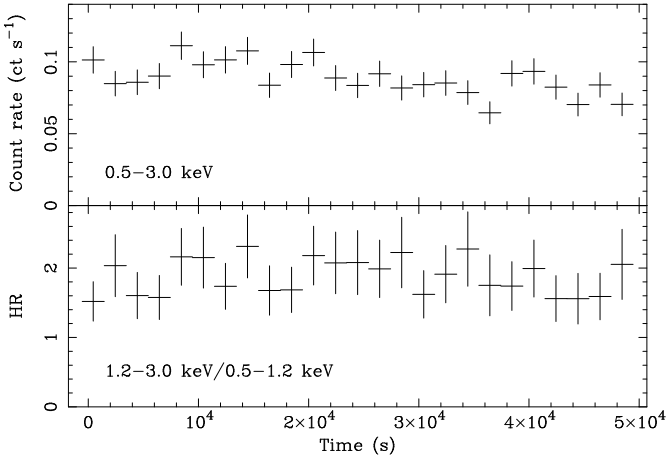}}
    \caption[]{\chandra\ ACIS I light curve and hardness ratio of \mx7 during
    observation 1730 integrated over 
    2000 s. Time zero  corresponds to HJD 245\,1737.88019 (solar 
     system barycenter corrected).  The
    light curves cover orbital phases 0.49--0.65 (ephemeris of 
    \citetalias{1999MNRAS.302..731D}).
    \label{ch_1730_lc}}
\end{figure}
\begin{figure*}
  \resizebox{6.cm}{!}{\includegraphics[bb=43 133 567 659,clip]{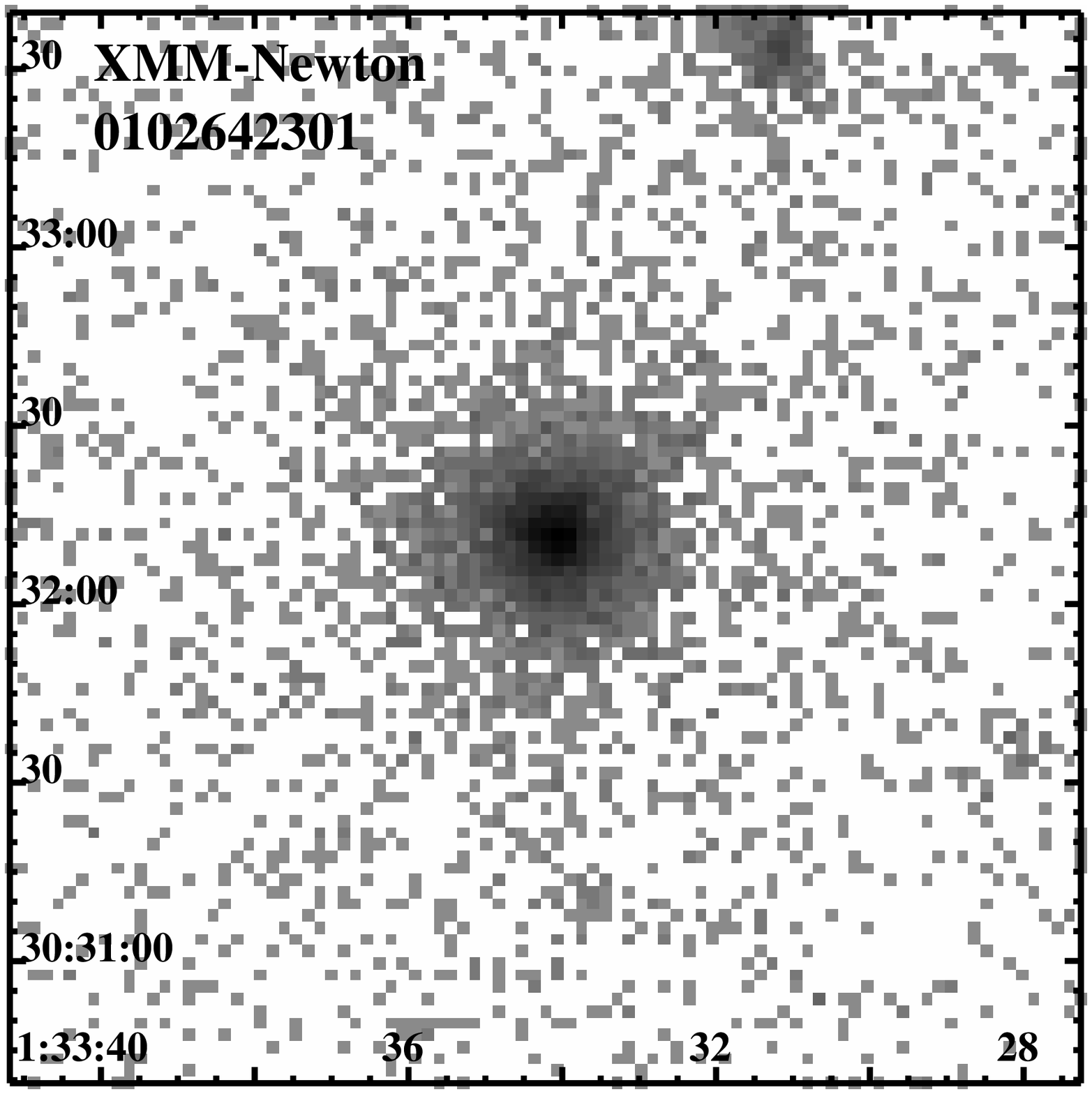}}
  \resizebox{6.cm}{!}{\includegraphics[bb=43 133 567 659,clip]{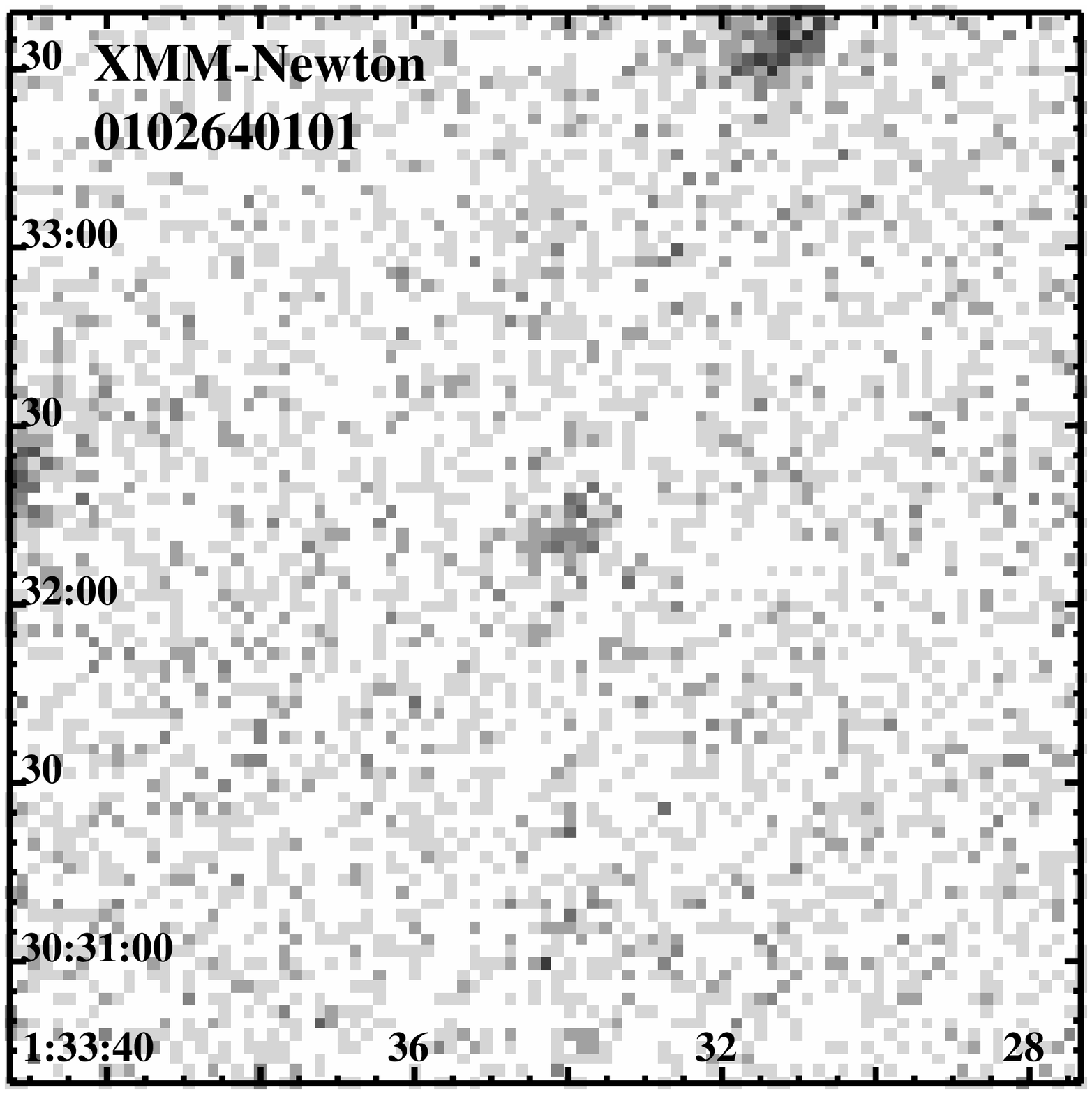}}
  \resizebox{6.cm}{!}{\includegraphics[bb=40 130 571 663,clip]{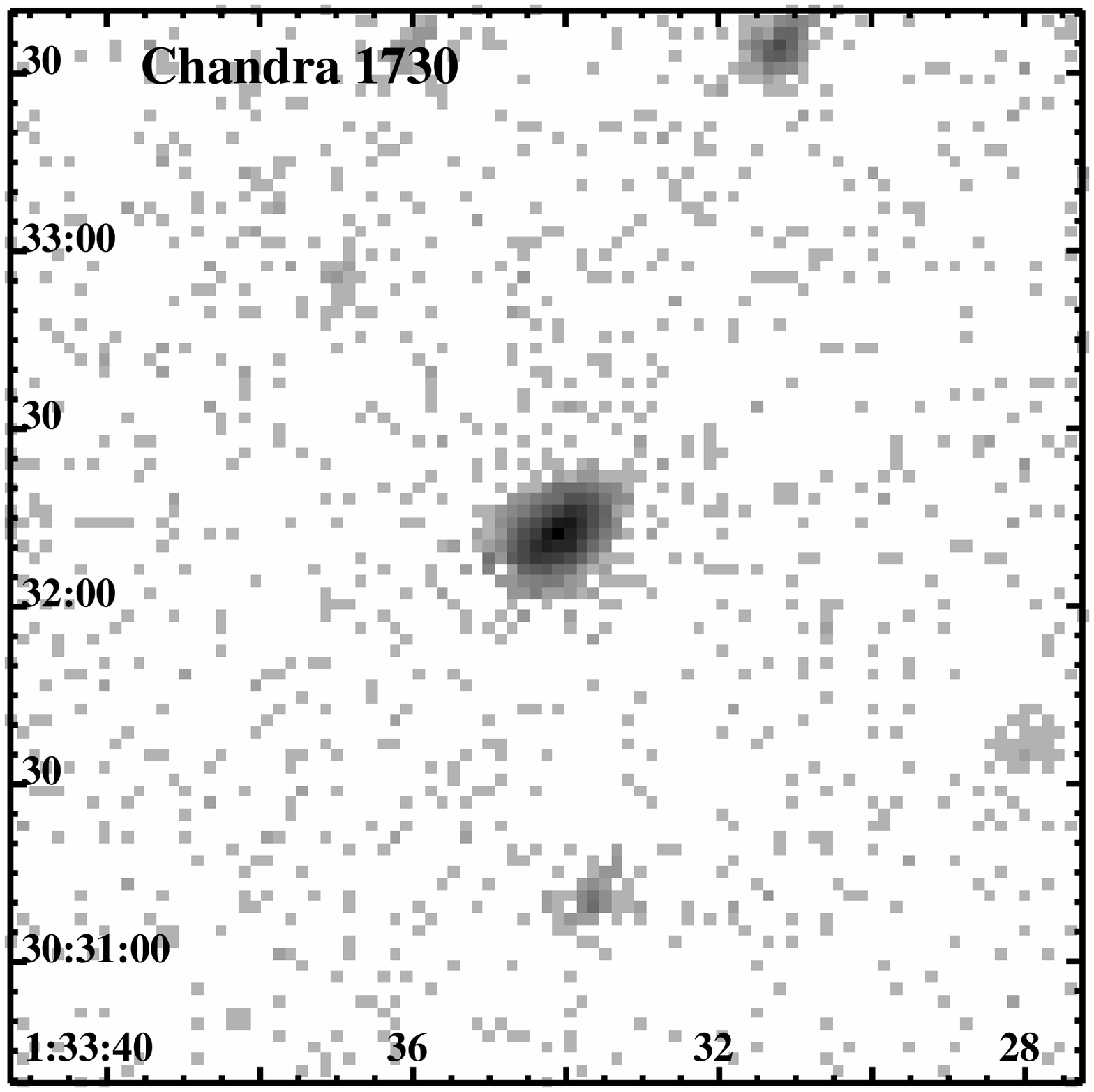}}
    \caption[]{
  Logarithmically-scaled, grey-scale images of the \mx7\ region of the 
  \xmm\ EPIC observations 0102642301, 0102640101, and
  \chandra\ ACIS I observation 1730 (from left to
  right).
  The images (RA, Dec J2000.0) were accumulated with a pixel size of 2\arcsec,
  in the 0.5--4.5 keV band. EPIC PN and MOS images were selected for times 
  of low background and added. Integration times for observation 
  0102642301 are 9.0, 12.1, and 12.2 ks for PN, MOS1, and MOS2, respectively,
  and the maximum in the image is 193 counts per 
  pixel. The corresponding numbers for observation 0102640101 are 11.7, 10.6,
  and 12.4 ks and 12 counts per pixel. In this image \mx7 is close to a CCD gap
  in PN and there is no MOS exposure in the upper left corner of the image.
  For the \chandra\ image the integration time was 49.4 ks and the maximum in
  the image is 919 counts per pixel.
  \mx7 is the source in the center of the images and was in a low state (eclipse)
  during observation 0102640101. HP01 X67 is the source $\sim1\farcm5$ to the 
  NNW of \mx7.  
  }
    \label{x_images}
\end{figure*}
\begin{figure*}
\begin{minipage}[t]{12cm}
\resizebox{\hsize}{!}{\includegraphics[bb=112 25 387 715,angle=-90,clip]{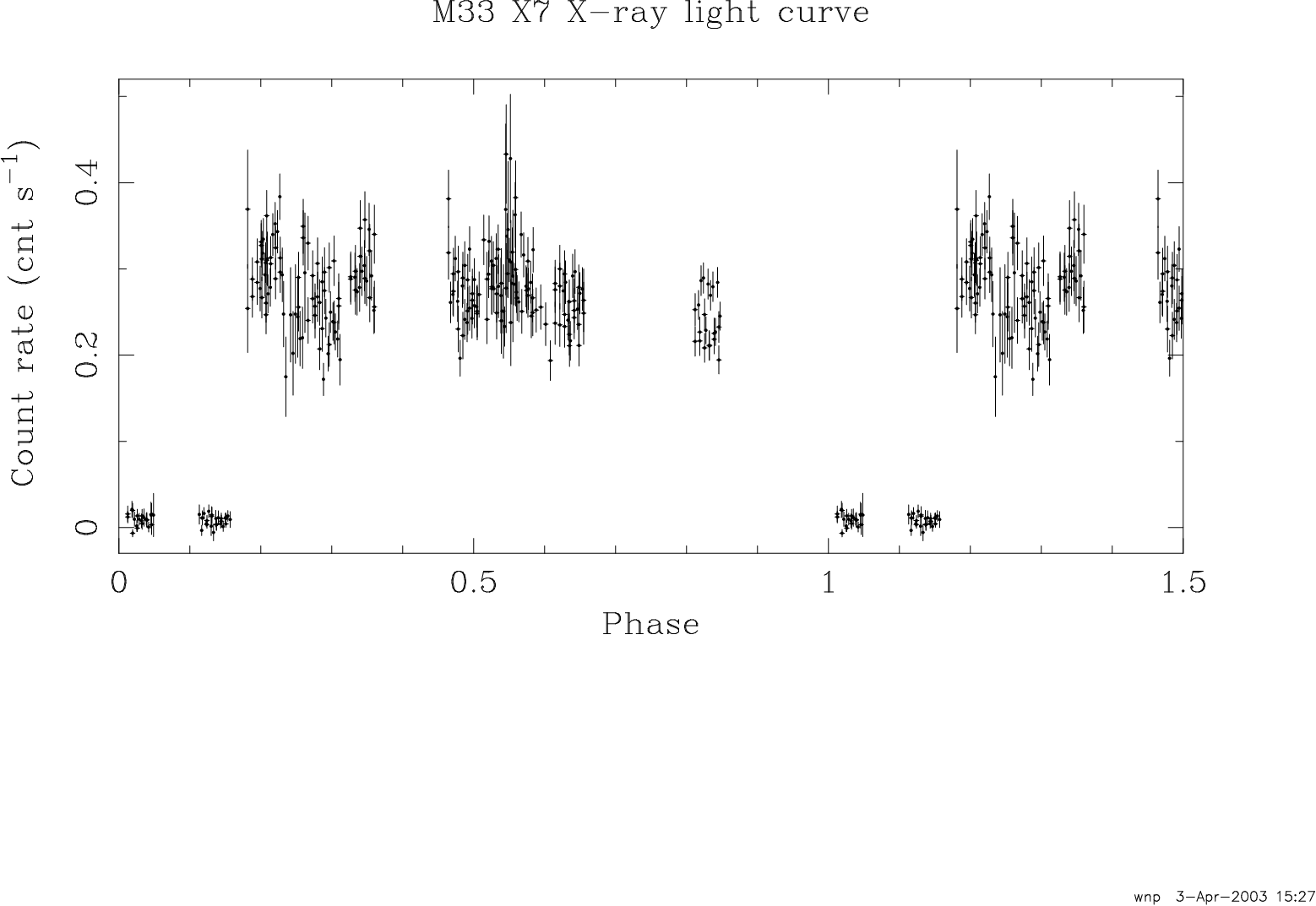}}
\resizebox{\hsize}{!}{\includegraphics[bb=80 25 549 715,angle=-90,clip]{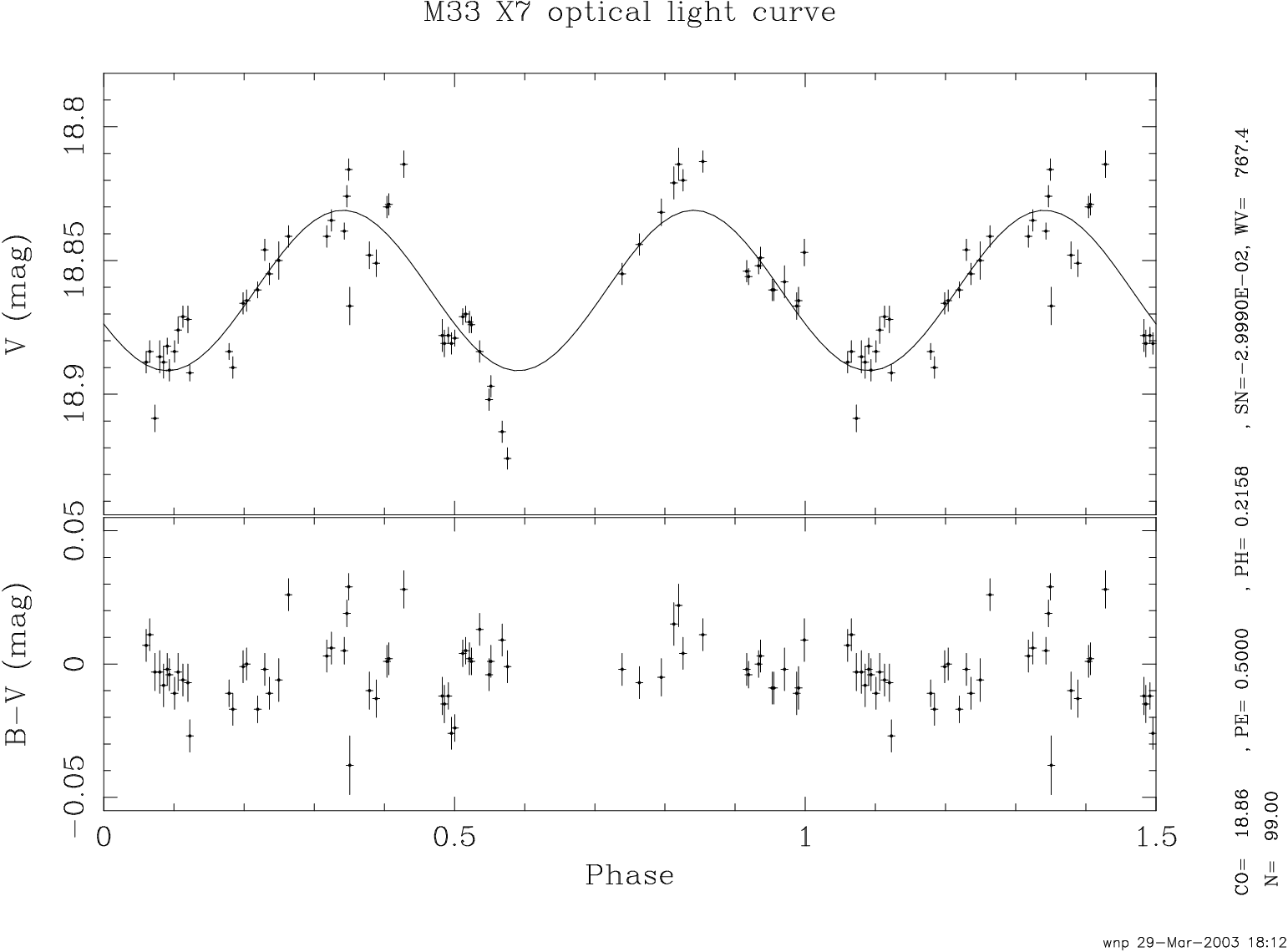}}
\end{minipage}
\begin{minipage}[t]{6cm}
\vspace{10cm}
\caption{Light curve of the XRB M33 X7 in the 0.5--3.0 keV band and in optical V
and B-V 
folded over the 3.45 d orbital period using ephemeris of 
\citetalias{1999MNRAS.302..731D}. Data for phase 0. to 0.5 are repeated at 1.0 to 1.5 for
clarity. Added is a double-sinusoidal approximation to the V-data (full line, 
    see text).}
\label{phase_x7}
\end{minipage}
\end{figure*}

The X-ray light curve of \x7 was sampled with a time resolution of 1000~s for
EPIC PN and with 2000~s for MOS. While we give raw count rate in 
Table~\ref{observations}, the light curves in Fig.~\ref{epic_lc} and 
\ref{phase_x7} are corrected for vignetting and EPIC MOS count rates for 
the difference in instrument efficiency compared to EPIC PN (i.e. increased by a 
factor of 2.5). The vignetting corrections for far off-axis positions is up to
a factor of 3.5 and may include systematic errors of up to 20\%.  
The source extraction radii chosen assure that more than 80\% of the PSF are
covered.
EPIC PN as well as MOS show \x7  at low intensity for \xmm\ observations
0102640101 and 0102641101 which covered binary phase 0.01 to 0.16 according to
the ephemeris of \citetalias{1999MNRAS.302..731D}. 
 During the other \xmm\ observations (covering binary phase 0.18 to 0.85 according to
the ephemeris of \citetalias{1999MNRAS.302..731D}, see Fig.~\ref{epic_lc} and 
\ref{phase_x7}), the 
source was in high state. The high state intensity did not vary by more than
$\pm20\%$ between the observations. Within the observations the intensities
changed by up to 30\% on time scales as short as 1000~s (see
Fig.~\ref{epic_lc}). No significant variability in hardness ratio was detected
between the observations. However, there seems to be variability on shorter
time scales. Some of the observations suffer from strong particle background 
flares which show vignetting effects by the telescope mirrors. 
We did not exclude these times from the light curves but tried to
correct by subtracting this variable background 
determined as close to \x7 as allowed by CCD configuration and emission from
other nearby sources. Our selection of the background regions should minimize
spurious effects on source count rate and hardness ratio. Nevertheless, 
residual effects may explain the count rate increase in observation 
0141980701 around binary phase 0.55 seen in Fig.~\ref{phase_x7}.

The light curve for the \chandra\ observation 1730 shows the source out of
eclipse. While the count rate in the 0.5--3.0 keV band decreased during the
observation by about 20\% and there is also variability down to the the sampling
time scale of 2000 s, the hardness ratio did not change significantly  
(Fig.~\ref{ch_1730_lc}).

One has to keep in mind that the EPIC MOS count rates and HRs 
of \x7 can not
directly be compared to the PN one as they originate from a CCD with a 
different energy response function. The same is even more true if one wants to
compare \chandra\ and \xmm\ count rates and HRs.

Figure~\ref{x_images} shows images (3\arcmin\ to a side) of the \x7 area in 
the 0.5--4.5 keV band during the high and low state ( \xmm\ EPIC cameras are
 combined).  
\x7 is clearly detected also during the low state ($\sim3\%$ of the average 
high state emission).

We searched for pulsations within the bright state \xmm\ low background
observations and the
\chandra\ observation in the frequency range 10$^{-3}$--4 Hz  following
the Rayleigh Z$^2_{\rm n}$ method \citep{1983A&A...128..245B} as described in
\citet{2002A&A...391..571H}. No signal was found with more than 94\%
($<2\sigma$) confidence level. This includes the previously suggested period of
0.31~s \citepalias{1999MNRAS.302..731D} although the time resolution of 0.073~s
of the PN detector in the full frame mode may not be sufficient to clearly 
detect such a short period.

\subsection{Improved position}
\begin{figure}
   \resizebox{\hsize}{!}{\includegraphics[]{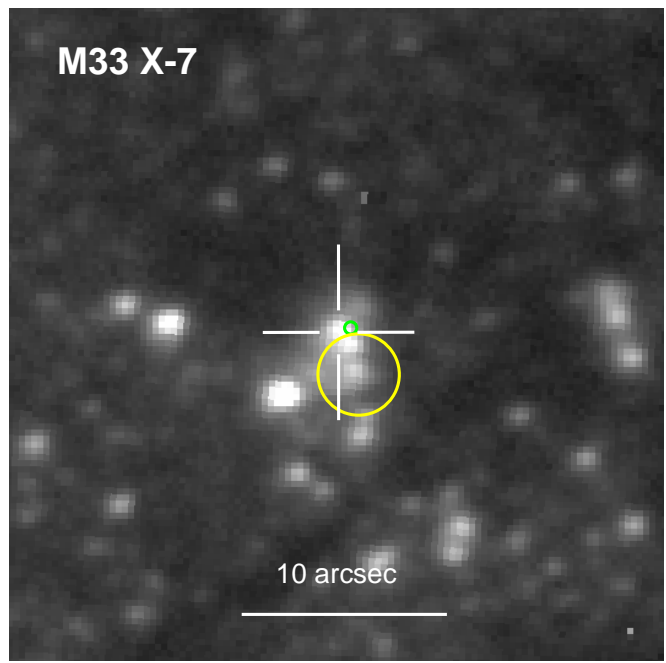}}
    \caption[]{Position of \mx7\ is shown on a V image of the DIRECT survey. The
    ROSAT position and error from \citetalias{2001A&A...373..438H} is marked by 
    the big circle, \chandra\  position by the small circle. 
    The optical identification, i.e. the star showing the 3.4~d 
    variability, is marked by the cross hair.
    \label{x7_opt}}
\end{figure}
\citetalias{2001A&A...373..438H} give the position of \x7\ with a 90\% error radius of 2\farcs0 which is 
fully
determined by the assumed remaining systematic error of 2\farcs0. 
The much higher number of photons detected with the \chandra\ ACIS I detectors
and the good PSF even at an off-axis angle of 8\farcm6 
allow us to determine a significantly improved source position in observation 
1730. Systematic
errors can be reduced by adjusting the positions using the well determined
radio position of a supernova remnant (SNR) close to \x7 
\citep[source 57 in ][]{1999ApJS..120..247G} which is also
detected in X-rays \citepalias[X67 in ][]{2001A&A...373..438H},  
the X-ray source $\sim1\farcm5$ to the NNW of \x7 in
Fig.~\ref{x_images}. We get a significantly improved position for \x7:
$\alpha$=01$^{\rm h}$33$^{\rm m}$34\fs21,
$\delta$=$+$30\degr32\arcmin11\farcs7 
(J2000), with a remaining error radius of 0\farcs3 mainly determined from the
statistical uncertainty of the X67 position.

Several \xmm\ EPIC observations cover \x7 and in addition X67. The integration
times of the individual observations are much shorter than for the \chandra\
observation and the \xmm\ PSF is worse. \xmm\ determined positions of \x7 have
individual error radii of 1\farcs0 or greater for the PN and MOS cameras 
and are within the errors consistent with the \chandra\ position. 
See Fig.~\ref{x7_opt} for an overlay of the improved position on a
deep optical $V$-band image.

\subsection{Energy spectra}
\begin{figure*}
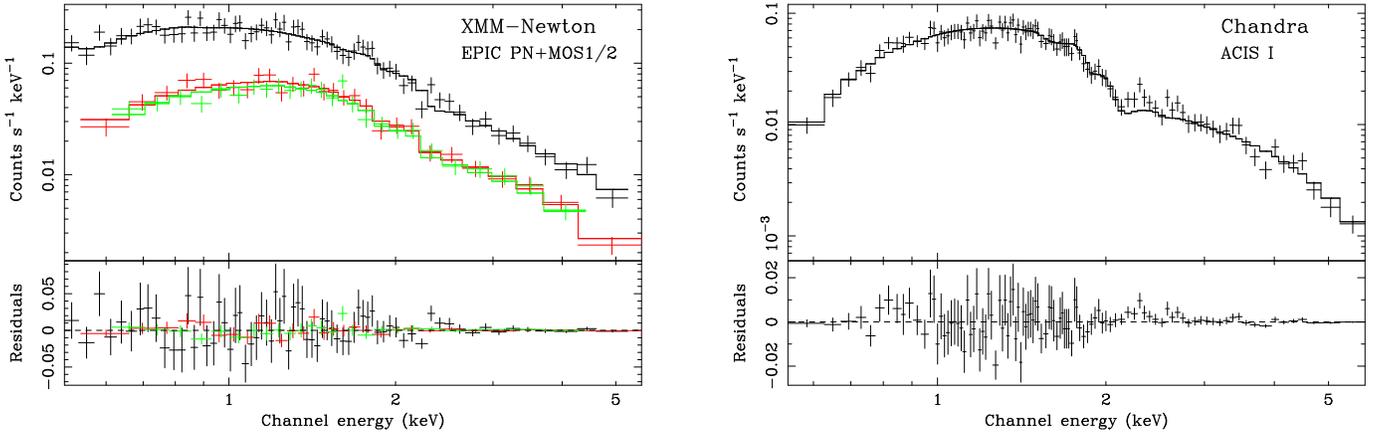

   \resizebox{8.5cm}{!}{\includegraphics[bb=100 30 562 700,angle=-90,clip]{0081f6a.ps}}
   \hspace{1.0cm}
   \resizebox{8.5cm}{!}{\includegraphics[bb=100 30 562 700,angle=-90,clip]{0081f6b.ps}}
    \caption[]{\mx7 spectrum of \xmm\ EPIC PN observation 0102641201 (left) and 
    of \chandra\ ACIS I observation 1730 (right). 
    Data and best fitting bremsstrahlung model are shown in the 
    upper panels, residuals between data and model below. 
    For the model parameters see Table~\ref{spectralfits}.
    \label{spec_x7}}
\end{figure*}

\begin{table*}
\begin{center}
\caption{Spectral modeling results for \mx7 for the on-axis \xmm\
EPIC observation 0102641201 (all EPIC instruments fitted together) 
and the \chandra\ ACIS I observation 1730. The degradation 
of the ACIS was taken into account using model {\sl ACISABS} in XSPEC.
For each instrument, we give the effective integration time $T_{\rm int}$ and
the raw count rate. The number of energy bins reduced by the number of free
parameters defines the degrees of freedom $\nu$.
In the case of $\chi^2/\nu\le2.0$, 90$\%$ errors are given. }
\label{spectralfits}

\begin{tabular}{llrrrlrrrrr}
\hline
\noalign{\smallskip}
\hline
\noalign{\smallskip}
Observation & Inst. & $T_{\rm int}$ & Raw count rate & $\nu$ & Model$^*$ 
& $N_{\rm H\,M33}^{**}$ & 
     $\Gamma$&k$T$  & $L_{\rm X}^{***}$ & $\chi^2/\nu$\\
&& (ks) &($10^{-2}$ ct s$^{-1}$)& & & ($10^{20}$~cm$^{-2}$) & & (keV) &
    ($10^{37}$ erg s$^{-1}$) \\
\noalign{\smallskip}\hline
\noalign{\smallskip}
\xmm &PN& 9.0 &32.0$\pm$0.6 & 119 & POWL  & $19.3_{-2.8}^{+3.1}$ &   
$2.24_{-0.09}^{+0.09}$&&  $6.9_{-0.6}^{+0.6}$  & 0.95\\ 
\noalign{\smallskip}   
EPIC &MOS1& 12.1&10.5$\pm$0.3 && BREMS & $8.8_{-2.0}^{+2.0}$ && $2.93_{-0.27}^{+0.32}$       
&  $7.0_{-0.5}^{+0.5}$ &0.79  \\\noalign{\smallskip} 
0102642301&MOS2&12.2&9.2$\pm$0.3 &&THPL  & $2.3_{-1.4}^{+1.5}$&&$4.7_{-0.2}^{+0.4}$ 
& $7.0_{-0.3}^{+0.3}$  & 1.47 \\ \noalign{\smallskip}
&&&&&DISKBB &$<0.8$& &$0.96_{-0.03}^{+0.04}$& $7.1_{-0.8}^{+1.0}$ 
&0.87\\
\noalign{\smallskip}
\noalign{\smallskip}
\noalign{\smallskip}
\chandra &ACIS I& 49.4 & 10.2$\pm$0.2 & 101 & POWL  & $28.4_{-3.8}^{+4.1}$ &  
$2.33_{-0.08}^{+0.11}$&&  $6.2_{-0.6}^{+0.6}$  &1.23\\
\noalign{\smallskip}
1730&&&&& BREMS & $15.3_{-2.6}^{+2.8}$ && $2.82_{-0.24}^{+0.27}$      
&  $6.3_{-0.4}^{+0.4}$ & 0.98 \\
\noalign{\smallskip}
&&&&&THPL  &      7.1     && 4.7           & 6.3  & 2.1 \\\noalign{\smallskip}
&&&&&DISKBB & $3.2_{-2.3}^{+2.4}$& &$1.01_{-0.05}^{+0.04}$& $6.5_{-1.1}^{+1.3}$ &0.97\\
\noalign{\smallskip}\hline
\noalign{\smallskip}
\end{tabular} 

\end{center}
$^*$\  : \hskip.3cm THPL = thin thermal Plasma with solar abundance (XSPEC model MEKAL), 
BREMS = thermal bremsstrahlung, \\
POWL = power law, DISKBB = disk blackbody\par
$^{**}$ : \hskip.3cm Absorption exceeding the fixed Galactic foreground\par
$^{***}$: \hskip.3cm In the 0.5--4.5 keV band, corrected for Galactic absorption,
corrected for extraction radii and vignetting\par
\end{table*}   

\begin{table*}
\begin{center}
\caption{Spectral modeling results for the remaining out of eclipse \xmm\
EPIC observations of \mx7 using thermal bremsstrahlung and disk
blackbody spectra. In the last row, five EPIC PN spectra of low background 
observations ($\Sigma$ EPIC PN; combining 0102640501, 0102640701, 0102642301,  
0141980601, and 0141980801) are simultaneously fitted.
Model parameters are given with 90$\%$ errors (see Table~\ref{spectralfits}). 
}
\label{epic_spectra}
\begin{tabular}{lrrrrrrr}
\hline
\noalign{\smallskip}
\hline
\noalign{\smallskip}
Observation & $\nu$& \multicolumn{3}{l}{Thermal bremsstrahlung} & 
 \multicolumn{3}{l}{Disk blackbody} \\
&&$N_{\rm H\,M33}$ &k$T$& $\chi^2/\nu$&$N_{\rm H\,M33}$ &k$T$& $\chi^2/\nu$\\
&& ($10^{20}$~cm$^{-2}$) & (keV) &&($10^{20}$~cm$^{-2}$) & (keV) &\\
\noalign{\smallskip}\hline
\noalign{\smallskip}
0102641201 & 53 &$16.0_{-3.3}^{+3.3}$ & $2.16_{-0.22}^{+0.26}$ & 0.79 
                 &$4.2_{-2.8}^{+3.1}$ & $0.87_{-0.05}^{+0.06}$ & 0.96\\
0102640401 & 68 &$12.3_{-3.0}^{+3.1}$ & $2.66_{-0.27}^{+0.33}$ & 1.05 
                 &$<2.6$ & $0.99_{-0.06}^{+0.04}$ & 1.20\\
0102640501 & 53 &$13.5_{-3.6}^{+3.8}$ & $2.72_{-0.33}^{+0.39}$ & 1.00 
                 &$<4.6$ & $0.99_{-0.06}^{+0.06}$ & 0.99\\
0102640601$^{+}$ & 18 &$11.4_{-9.3}^{+10.4}$ & $2.87_{-1.26}^{+4.29}$ & 0.88 
                 &$<11.7$ & $0.90_{-0.23}^{+0.25}$ & 0.99\\
0102640701 & 41 &$10.8_{-4.0}^{+4.3}$ & $2.90_{-0.47}^{+0.61}$ & 0.91
                 &$<3.3$ & $0.98_{-0.07}^{+0.06}$ & 0.93\\
0102642101$^{*}$ & 30 &$14.4_{-4.6}^{+5.1}$ & $3.00_{-0.61}^{+0.87}$ & 0.97
                 &$<8.4$ & $0.97_{-0.10}^{+0.11}$ & 0.97\\
0141980501 & 45 &$12.7_{-4.3}^{+4.6}$ & $2.54_{-0.39}^{+0.48}$ & 0.61
                 &$<4.8$ & $0.94_{-0.08}^{+0.07}$ & 0.79\\
0141980601 & 75 &$17.0_{-2.7}^{+2.9}$ & $2.14_{-0.18}^{+0.21}$ & 1.25
                 &$4.8_{-2.4}^{+2.6}$ & $0.88_{-0.05}^{+0.05}$ & 1.15\\
0141980701 & 87 &$9.4_{-3.9}^{+4.1}$ & $2.83_{-0.58}^{+0.89}$ & 1.06
                 &$<3.9$ & $0.90_{-0.09}^{+0.08}$ & 1.04\\
0141980801 & 80 &$14.7_{-2.8}^{+3.9}$ & $2.13_{-0.17}^{+0.20}$ & 1.01
                 &$<4.4$ & $0.89_{-0.05}^{+0.04}$ & 0.92\\
\noalign{\smallskip}		 
$\Sigma$ EPIC PN& 251 &$11.6_{-1.5}^{+1.6}$ & $2.47_{-0.15}^{+0.17}$ & 1.07
                 &$<1.7$ & $0.94_{-0.03}^{+0.03}$ & 1.04\\
\noalign{\smallskip}\hline
\noalign{\smallskip}
\end{tabular} 
\end{center}
Notes and references:\\
$^{ +~}$: \mx7 only in PN field of view. \\
$^{ *~}$: \mx7 only in MOS field of view. \\
\end{table*}

Spectra for the low background \xmm\ EPIC observation 0102642301 and for the 
\chandra\ observation were approximated with simple spectral models 
(see Table~\ref{spectralfits}).
Individual EPIC PN and MOS spectra were simultaneously fitted with the same 
model parameters correcting for vignetting and the fraction of the PSF covered
by the source extraction area.
Energy independent normalization factors for each spectrum
separately take 
into account possible differences in source coverage. Model components were  
the fixed Galactic foreground absorption  
plus additional absorption within \m33 or in the immediate surrounding of the
source using XSPEC model component WABS.
A thin thermal (MEKAL) model could be rejected due to the high 
$\chi^2/\nu$, and also a power law (POWL) model still shows large residuals. 
Bremsstrahlung (BREMS) and also  
variable temperature disk blackbody (DISKBB) models on the other hand yield 
acceptable $\chi^2/\nu$ values. We used the latter models
to fit all high state observations of \x7 (see Table~\ref{epic_spectra}).
For all observations, absorbed BREMS or DISKBB models give acceptable fits.
The BREMS model needs comparably high absorption outside the Galaxy. The 
model temperatures are the same within the errors. For the DISKBB model, 
on the other hand, the additional
absorption is very low or absent and inner disk temperatures also don't vary
significantly.

\section{Optical observations and results}
\subsection{Observations}
The data were obtained as part of the DIRECT project\footnote{For
information on the DIRECT project see {\tt
http://cfa-www.harvard.edu/\~\/kstanek/DIRECT/}.} 
\citep[][ \citetalias{2001AJ....122.2477M}]{2001AJ....121.2032M,2001AJ....121.3284M}
at the Kitt Peak National Observatory\footnote{Kitt Peak
National Observatory is a division of NOAO, which are operated by the
Association of Universities for Research in Astronomy, Inc. under
cooperative agreement with the National Science Foundation.} 2.1m
telescope during two separate runs, from September 29th to October 5th,
1999 and from November 1st to 7th, 1999. The telescope was equipped with
a Tektronix $2048\times2048$ CCD (T2KA camera) having a pixel scale
$0.305\arcsec/pixel$. We collected $74\times600$~s exposures in the $V$
filter and $30\times600$ s in $B$. 
The exposure times varied slightly to compensate for the changes
of seeing conditions. The typical seeing was 1\farcs5. 

\subsection{Data reduction and calibration}
The preliminary processing of the CCD frames was performed with the
standard routines in the IRAF ccdproc package.\footnote{IRAF is
distributed by the National Optical Astronomy Observatories, which are
operated by the Association of Universities for Research in Astronomy,
Inc., under cooperative agreement with the NSF.} The data were corrected
for CCD non-linearity at this stage, as described by 
\citet{2001AJ....121.2032M}.

Photometry was extracted using the ISIS image subtraction package
\citep{1998ApJ...503..325A,2000A&AS..144..363A}. 
A brief outline of the method is presented
here. For a more detailed description of the reduction procedure the
reader is referred to \citet{2001AJ....121.2032M}.

The ISIS reduction procedure consists of the following steps: (1)
transformation of all frames to a common $(x,y)$ coordinate grid; (2)
construction of a reference image from several best exposures; (3)
subtraction of each frame from the reference image and (4) extraction of
profile photometry from the subtracted images.

All computations were performed with the frames internally subdivided
into four sections ({\tt sub\_x=sub\_y=2}). Differential brightness
variations of the background were fit with a second degree polynomial
({\tt deg\_bg=2}). A convolution kernel varying quadratically with
position was used ({\tt deg\_spatial=2}). The PSF width ({\tt
psf\_width}) was set to 15 pixels. We used a photometric radius ({\tt
radphot}) of 3 pixels.

Due to residual non-linearity, our photometry could not be calibrated
from observations of standard stars. The coefficients for the color
terms of the transformation were derived from the comparison of our NGC
6791 photometry with the data from the KPNO 0.9m telescope 
\citep{1992AcA....42...29K}.  
The offsets were determined relative to 735 stars above
$V=20$ mag from the DIRECT catalog of stellar objects in M33 
\citep{2001AJ....121..861M}.
The calibration coefficients can be found in \citetalias{2001AJ....122.2477M}.

The transformation from rectangular to equatorial coordinates was
derived using 894 transformation stars with $V<19.5$ from the DIRECT
catalog of stellar objects in M33 \citep{2001AJ....121..861M}. The average
difference between the catalog and the computed coordinates for the
transformation stars was 0\farcs06 in right ascension and
0\farcs06 in declination.

\subsection{Results}
\begin{table}
\begin{center}
\caption[]{Optical light curve of the star identified with \mx7. Magnitude with error in the V or B filter
(column F) and heliocentric Julian date (HJD) of the observation are given. }
\begin{tabular}{rrc|rrc}
\hline\noalign{\smallskip}
\hline\noalign{\smallskip}
\multicolumn{1}{c}{F} & 
\multicolumn{1}{l}{HJD $-$} & 
Mag &
\multicolumn{1}{c}{F} & 
\multicolumn{1}{l}{HJD$-$} & 
Mag \\
&2451400&&&2451400 &\\
\noalign{\smallskip}
\hline\noalign{\smallskip}
V & 52.721458 &  18.854(4) &B & 56.793912 &  18.887(4) \\ 
V & 52.730637 &  18.856(3) &V & 56.811296 &  18.884(4) \\  
V & 52.780012 &  18.852(3) &B & 56.820741 &  18.873(4) \\  
V & 52.788819 &  18.849(4) &V & 56.829560 &  18.876(5) \\  
B & 52.798183 &  18.852(3) &B & 56.840301 &  18.865(4) \\  
V & 52.846146 &  18.861(4) &V & 56.853021 &  18.871(4) \\  
V & 52.854931 &  18.861(4) &V & 56.877454 &  18.872(5) \\  
V & 52.907662 &  18.858(6) &V & 56.887141 &  18.892(3) \\ 
B & 52.916632 &  18.856(5) &V & 57.674028 &  18.867(7) \\  
V & 52.967407 &  18.867(5) &B & 57.699340 &  18.829(8) \\  
V & 52.976493 &  18.865(5) &V & 57.771528 &  18.848(5) \\  
V & 53.005012 &  18.847(5) &V & 57.804664 &  18.851(5) \\  
V & 53.919248 &  18.841(4) &B & 57.814074 &  18.838(4) \\  
V & 54.676273 &  18.878(6) &V & 57.857789 &  18.830(4) \\  
V & 54.686343 &  18.881(5) &V & 57.866620 &  18.829(4) \\  
B & 54.696771 &  18.866(3) &B & 57.875891 &  18.831(4) \\  
V & 54.705671 &  18.878(3) &V & 57.940764 &  18.814(5) \\  
V & 54.720822 &  18.881(4) &B & 57.950278 &  18.842(4) \\  
B & 54.729942 &  18.855(3) &B & 84.699097 &  18.873(3) \\  
V & 54.738519 &  18.879(3) &V & 84.708727 &  18.884(3) \\  
V & 54.777199 &  18.871(3) &V & 84.726852 &  18.890(4) \\  
V & 54.791528 &  18.870(3) &V & 84.777234 &  18.866(4) \\  
B & 54.800428 &  18.875(3) &B & 84.785775 &  18.865(4) \\  
V & 54.809769 &  18.873(4) &V & 84.794618 &  18.865(4) \\  
V & 54.819468 &  18.874(3) &V & 84.849259 &  18.861(3) \\  
V & 54.860521 &  18.884(4) &V & 84.883831 &  18.846(4) \\  
B & 54.869780 &  18.897(4) &B & 84.895440 &  18.844(3) \\  
B & 54.878403 &  18.898(4) &V & 84.906539 &  18.855(4) \\  
V & 54.906435 &  18.902(4) &V & 84.952697 &  18.850(7) \\  
V & 54.915324 &  18.897(4) &B & 86.634097 &  18.853(3) \\  
V & 54.971019 &  18.914(4) &V & 86.643438 &  18.855(4) \\  
B & 54.987986 &  18.923(4) &V & 86.727384 &  18.844(4) \\  
V & 54.997303 &  18.924(4) &B & 86.742824 &  18.837(4) \\  
V & 55.753866 &  18.832(5) &V & 86.897975 &  18.821(6) \\  
B & 55.762789 &  18.827(4) &V & 86.921400 &  18.814(6) \\  
V & 55.861551 &  18.820(4) &V & 88.643009 &  18.841(4) \\  
B & 55.870822 &  18.824(3) &B & 88.652106 &  18.843(3) \\  
V & 55.958808 &  18.813(4) &V & 88.664965 &  18.835(4) \\  
V & 56.672176 &  18.888(4) &V & 88.728738 &  18.839(3) \\  
B & 56.680891 &  18.895(3) &V & 88.741968 &  18.826(4) \\  
V & 56.690012 &  18.884(4) &V & 88.750914 &  18.816(4) \\  
B & 56.701400 &  18.883(3) &B & 88.760104 &  18.845(2) \\  
V & 56.715382 &  18.909(5) &B & 88.780637 &  18.824(3) \\  
B & 56.724583 &  18.906(4) &V & 88.878808 &  18.826(6) \\  
V & 56.739606 &  18.886(6) &V & 88.893958 &  18.830(7) \\  
B & 56.748368 &  18.883(5) &V & 88.927685 &  18.813(7) \\  
V & 56.757824 &  18.888(6) &V & 89.698310 &  18.912(4) \\  
B & 56.766597 &  18.880(4) &V & 89.710938 &  18.912(3) \\  
V & 56.775752 &  18.882(3) &V & 90.841655 &  18.838(5) \\  
V & 56.786528 &  18.891(4) &V & 90.886076 &  18.835(6) \\
\noalign{\smallskip}				     
\hline
\noalign{\smallskip}
\end{tabular}
\label{optmag}
\end{center}
\end{table}
We have searched within a radius of 36\arcsec\ around the SIMBAD position of
\x7 for stars with a period of around 3.45~d. From the 715 stars examined only
one was a binary with a period of about 3.46-3.48 days.
It is a star at position
$\alpha$=01$^{\rm h}$33$^{\rm m}$34\fs20, $\delta$=$+$30\degr32\arcmin11\farcs08 
(J2000),
with an average V magnitude of 18.86, and an amplitude of variation 
in this band of about 0.11 mag.
From the images it seems to be located in a small
star cluster (see Fig.~\ref{x7_opt}). Measured V and B magnitudes are given in
Table~\ref{optmag}.

The binary light curve in the V band 
using the ephemeris of \citetalias{1999MNRAS.302..731D} (Fig.~\ref{phase_x7}) shows two maxima and minima of
different depth as expected for an ellipsoidal and X-ray heating light curve
for a high mass XRB (HMXB) 
together with a double-sinusoidal fit to the data (average V magnitude of 
$18.862\pm0.001$, amplitude $(33\pm1)\times 10^{-3}$ mag). The phase of the
shallower minimum is shifted by $+0.0956\pm0.0028$ with respect to the eclipse 
center of \citetalias{1999MNRAS.302..731D} in the same direction as also 
indicated by the X-ray light curve (see Sect. 2.1). The double-sinusoidal fit
is only a crude approximation to the data and the model parameters certainly are
affected by non-uniform sampling of the optical light curve. The data suggest
a deeper secondary minimum (around phase 0.6) which according to 
\citet{1986A&A...154...77T} might point at the effect of an accretion disk in
the system.
V magnitude and B--V color at X-ray eclipse are 18.890 and -0.005. 

\section{Discussion}
\subsection{Improved ephemeris}
\citetalias{1999MNRAS.302..731D} determined the eclipse
parameters for \x7\ using in total 50 ks and 325 ks of archival ROSAT PSPC and
HRI data, respectively, integrated over one satellite orbit ($\sim$3 ks time
bins). 
The \xmm\ and \chandra\ low and high 
state observations of \x7\ can naturally be explained by finding the source 
within and out of eclipse. Unfortunately, neither of the 
observations covered eclipse ingress or egress. However, 
from the \xmm\ light curves we can deduce that the eclipse egress lasts $<0.02$ and the 
ingress $<0.16$ in phase (well within the parameter range given by 
\citetalias{1999MNRAS.302..731D}). These values are only poorly determined due 
to gaps in the phase sampling over the orbit. More observations are needed to 
further constrain the binary ephemeris. 

Nevertheless, we can  
restrict the time of eclipse egress to HJD~($245\,1760.935\pm0.035$)
assuming that it is constrained by the end of observation 0102640101 
(when \x7 is still in eclipse) and the beginning of observation 0141980501 
(when, 261 binary orbits later, \x7 is already out of eclipse). With the
eclipse shape parameters of \citetalias{1999MNRAS.302..731D} we then derive a mid-eclipse
epoch of HJD~($245\,1760.61\pm0.09$). 
The new X-ray mid-eclipse time corresponds to binary phase 0.07 in
Fig.~\ref{phase_x7}. 
Combining this result with the mid-eclipse epoch of 
\citetalias{1999MNRAS.302..731D} we get an improved orbital period for \x7\ of
($3.45376\pm0.00021$) d well within the period errors given by 
\citetalias{1999MNRAS.302..731D}. 

\subsection{The optical companion}
The spectral classification of the optical companion can be deduced from 
the absolute 
optical magnitude and color during X-ray eclipse when we see the optical 
surface that is mostly undisturbed by gravitational effects, an expected 
accretion disk and heating by the X-ray source. 
To derive the absolute magnitude the measured brightness has to be 
corrected for the distance (-24.502 mag for the assumed distance of 795 kpc) 
and for interstellar extinction, the color has to be corrected for reddening.
These corrections can be estimated from the Galactic \nh\ in the direction of \x7
as $A_{\rm V\,Gal} = 0.36$ mag and $E(B-V)_{\rm Gal} = -0.12$ 
\citep{1995A&A...293..889P}.
The \nh\ column depth of the M33 disk in the direction of \x7\ is varying
significantly on small scales. \x7\ is not located in one of the 
\ion{H}{i} holes catalogued by \citet{1990A&A...229..362D}.
The absorbing column within M33 can be determined to $\sim$2.2\hcm{21} from a
$47\times 93$ arcsec half power beam width \ion{H}{i} map 
\citep{1980MNRAS.191..615N}. 
The absorbing columns derived from the spectral fits to the X-ray spectra
indicate half or less than this absorbing column depending on the spectral model.
Assuming that we see \x7\ through half of this column and that the same
conversion applies as used above,  we get $A_{\rm V\,M33} = 0.62$ mag and
$E(B-V)_{\rm M33} = -0.21$. 
The companion star therefore should have an absolute V magnitude of -6.0 to 
-6.6 and (B-V)$_0$ of -0.12 to -0.33, which would correspond to a star of
spectral typ  B0I to O7I and masses $M_{2}$ of 25 to 35 M$_{\sun}$ \citep[see
e.g.][]{1982lbor.book.....A}. 
Optical companions of similar spectral type were
proposed by \citetalias{1999MNRAS.302..731D} based on binary orbit and eclipse
length, assuming a 1.4 M$_{\sun}$ neutron star as the compact object. 

\subsection{\mx7, an eclipsing black hole XRB?}
We can use the mass of the optical companion as determined in the previous
subsection and the mass ratios determined from binary orbit and eclipse
length \citep[see \citetalias{1999MNRAS.302..731D} 
and][]{1971ARA&A...9..183P,1983ApJ...268..368E} to determine the  mass 
$M_{\rm X}$ of the compact object. For a binary inclination of 90\degr, 
the mass ratio q($=M_{\rm X}/M_{2}$) is 0.085 yielding $M_{\rm X}$ of 2.1--3.0
M$_{\sun}$, significantly higher than expected masses of neutron stars. 
For an inclination of 70\degr, q is 0.036 and $M_{\rm X} = 0.9-1.2$ M$_{\sun}$ 
well in the range of typical neutron star masses. The inclination can not
be constrained by our measurements and the mass estimates above do not exclude 
a neutron star for the compact object in the system.

Further information on the nature of the compact object may be infered from the
X-ray spectrum. Before \xmm\ and \chandra, the X-ray spectrum of \x7\ was 
investigated using ROSAT and \sax\ observations.   
\citetalias{2001A&A...373..438H} found that the \x7\ spectrum in the ROSAT
band can be described by an absorbed power law (\nh\ = (1.9$\pm$0.9)\hcm{21}
and $\Gamma = 1.86\pm0.52$). Thin thermal plasma models did not give acceptable
fits. \citet{2001A&A...368..420P} reported that \sax\ detected \x7\ in the 2--8
keV band. The spectrum could be described equally well by a power law 
(\nh\  $<$ 9.8\hcm{22} and photon index 
$\Gamma = 2.9^{+1.7}_{-1.3}$) or a Bremsstrahlung model
(\nh\  $<$ 11\hcm{22} and k$T = 3.7^{+97}_{-2.4}$). If the absorption is fixed
to 5.6\hcm{20}, the power law photon index is constrained to 1.7$\pm$0.6 and the
Bremsstrahlung temperature to $<$3.7 keV. Keeping in mind the large uncertainties
in the parameters, these results are consistent with the \xmm\ and \chandra\
spectra reported for this persistent source in Sect. 2.3. 
In the following we therefore concentrate on
these higher quality spectral results.   

Thermal Bremsstrahlung and disk blackbody models yield acceptable fits to the 
\xmm\ and \chandra\ spectra (see Sect. 2.3). For the first model 
an absorbing column comparable to half  the \ion{H}{i}  values at the position of
\x7\ within \m33\  (see above) is needed to fit the \x7\ spectra. The second 
model only allows us to determine upper limits for the \m33\ absorbing column in
front of \x7\ which are smaller than half the M33 column depth at that position.
The \x7\ X-ray spectra are steeper in the 2--10 keV band than typical for HMXB 
\citep[flat power law shapes with photon indices of $0.8 < \Gamma < 1.5$, ][]
{1983ApJ...270..711W} and better resemble the spectra of persistent black 
hole XRBs like \object{LMC X$-$1} and \object{Cyg X$-$1} \citep[see e.g. ][]{2002ApJ...576..357C}.

Therefore, the spectral results, our unsuccessful search for X-ray pulsations 
and the high mass estimate of the compact object,  infered from the 
identified optical companion indicate a black hole as the compact object. 
These findings make \mx7\ a very interesting source for further investigation as
it could be the first eclipsing HMXB with a black hole as the compact object.

\section{Conclusions}
\xmm\ and \chandra\ observations of the persistent eclipsing HMXB \mx7\ allowed us to
improve on the orbital period and investigate the X-ray spectrum in
unprecedented detail. No X-ray pulsations were detected. A special 
investigation of the optical variability of DIRECT data of the region revealed
in the optical the orbital light curve of a high mass companion. X-ray and
optical data point at a black hole as the compact object in the system.
Optical spectroscopy and high sensitivity X-ray pulsation searches are needed to
clarify the situation.

\begin{acknowledgements}
    The \xmm\ project is supported by the Bundesministerium f\"{u}r
    Bildung und Forschung / Deutsches Zentrum f\"{u}r Luft- und Raumfahrt 
    (BMBF/DLR), the Max-Planck Society and the Heidenhain-Stiftung.
\end{acknowledgements}
\bibliographystyle{apj}
\bibliography{./0081,/home/wnp/data1/papers/my1990,/home/wnp/data1/papers/my2000,/home/wnp/data1/papers/my2001}

\end{document}